\documentclass{article}

\usepackage{graphicx}
\newcommand{\Pd}{P^\dagger_{\sigma}}

\newcommand{\Pl}{P^{\vphantom{\dagger}}_{\sigma}}

\newcommand{\ad}{d^\dagger_{\sigma}}

\newcommand{\al}{d^{\vphantom{\dagger}}_{\sigma}}
\newcommand{\pmx}{p^{\vphantom{\dagger}}_{-\hat{x},\sigma}}
\newcommand{\pmy}{p^{\vphantom{\dagger}}_{-\hat{y},\sigma}}
\newcommand{\px}{p^{\vphantom{\dagger}}_{+\hat{x},\sigma}}
\newcommand{\py}{p^{\vphantom{\dagger}}_{+\hat{y},\sigma}}
\newcommand{\pdmx}{p^{\dagger}_{-\hat{x},\sigma}}
\newcommand{\pdmy}{p^{\dagger}_{-\hat{y},\sigma}}
\newcommand{\pdx}{p^{\dagger}_{+\hat{x},\sigma}}

\newcommand{\updw}[2]{\mbox{$\left|#1#2\right>$}}

\newcommand{\updo}[1]{\mbox{$\left|#1\right>$}}
\newcommand{\updwa}[2]{\mbox{$\left|#1\!\!\uparrow\right>\left|#2\!\!\downarrow\right>$}}
\newcommand{\dwupa}[2]{\mbox{$\left|#1\!\!\downarrow\right>\left|#2\!\!\uparrow\right>$}}

\title{Destabilization of the Zhang-Rice singlet at optimal doping}

\author{D.~K.~Sunko\\Department of Physics, Faculty of Science, University of
Zagreb,\\ Bijeni\v cka cesta 32, HR-10000 Zagreb, Croatia.}

\begin{document}

\maketitle
\begin{abstract}

The construction of the Zhang-Rice singlet is revisited in the light of recent
understanding of high-temperature superconductors at optimal doping. A minimal
local model is derived which contains the physical regime found relevant for
ARPES experiments, characterized by significant direct oxygen-oxygen hopping.
For the values of orbital parameters indicated by experiment, the Zhang-Rice
singlet is strongly mixed with a pure oxygen singlet of the same symmetry. The
destabilization of the Zhang-Rice ground state is due to the oxygen singlet
having twice as large a coherence factor with respect to oxygen-oxygen
hopping. An analogous quantum phase transition is identified in the
$t$--$t'$--$J$ model. The orbital-antisymmetric copper-oxygen singlet is
confirmed to be irrelevant, as found originally. The usual symmetry analysis
is extended to include dynamical symmetries. \end{abstract}


\section{Introduction}

The nature of the electronic wave functions which undergo the transition to
superconductivity in copper-oxide perovskites is still unknown. It is
generally at least tacitly assumed that the superconducting (SC) transition at
optimal doping does not invoke radically different states than those already
present at the Fermi level slightly above T$_c$. In this approach, shared
here, the understanding of the normal state is considered the key to the SC
mechanism.

The antiferromagnetic (AF) instability in the vicinity of half-filling in the
same compounds is by contrast well understood~\cite{Eskes93} as a
charge-transfer effect, dominating the first (large Mott-Hubbard interaction
$U_d$) Coulomb correction~\cite{Anderson59} to the ideal (paramagnetic) Mott
insulator~\cite{Mott49}. Theoretical attempts to connect this underdoped
region with the SC region on the hole-doped side abound in the literature.
They have received early support in the ionic limit by the
Zhang-Rice~\cite{Zhang88} construction, which considered an added hole in an
AF background, when double occupation of the copper site was blocked by a
large $U_d$. It is known that the hole then occupies the oxygen sites, each
bridging two coppers in the CuO$_2$ plane. Assuming that the hole prefers to
bind with one of the coppers, Zhang and Rice have shown that a uniform
spreading of the hole around that copper, in an orbitally symmetric
``Zhang-Rice singlet'' (ZRS) is energetically strongly preferred, and
propagates like a projected fermion. This provided a way to vindicate the
single-band $t$--$J$ model as a physical reduction of the Emery (three-band
Hubbard) model~\cite{Emery87,Varma87}, at least for underdoped systems, as an
alternative to the usual formal derivation, which applies the Foldy-Wouthuysen
transformation to the one-band Hubbard model~\cite{Fulde93}.

Challenges to this picture come from two directions. One is fundamental, that
there is no \emph{a priori} reason for the hole to prefer one copper over the
other. When it is undecided, triplet states are strongly admixed, while it may
still move about as a fermion~\cite{Emery88}. The other is to accept the basic
idea of a single CuO$_4$ molecule, but consider additional states and overlaps
within it, to find the limits of the particular construction
employed~\cite{Macridin05}. In the present work, the natural next step in the
Zhang-Rice construction is taken, which introduces just one additional
singlet, occupying the oxygen sites.

As will be shown below, such an extension is the minimal model to express the
low-energy physics of optimally doped high-T$_c$ superconductors in the
small-cluster language.  At optimal doping, the Zhang-Rice ground state
becomes strongly admixed with said oxygen singlet, chosen by the same symmetry
arguments as originally used by Zhang and Rice~\cite{Zhang88}.

The simplicity of the calculation is used as an opportunity to introduce
classification of state vectors by dynamical symmetry on a small example. This
method may have some potential for larger problems.

\section{Physical scales}

Quantum chemical calculations on small clusters predict the regime
\begin{equation}
U_d\gg\Delta_{pd}>|t_{pd}|\gg |t_{pp}|,
\label{qcr}
\end{equation}
used by Zhang and Rice, and confirmed in XPS experiments~\cite{Veenendaal94}.
Here $\Delta_{pd}=\varepsilon_p-\varepsilon_d$ is the copper-oxygen splitting,
positive in the hole picture, while $t_{pd}$ and $t_{pp}$ are the
copper-oxygen and oxygen-oxygen overlaps, respectively. In the limit
(\ref{qcr}), the ZRS is the ground state. The energy scale of the XPS
experiments and corresponding calculations is 5--10~eV. In high-T$_c$
materials, there is a universally acknowledged crossover scale of the order of
0.1~eV between the lowest and higher energies, although it is still debated
whether its origin is phononic~\cite{Lee07} or magnetic~\cite{Valla07}.
Neither small-cluster nor LDA calculations obtain this crossover, and the
latter are consistent with the high-energy result (\ref{qcr}). A recent
analysis~\cite{Meevasana07} showed three distinct many-body scales in BSCCO
which were not reproduced by contemporary LDA calculations, the lowest of
which was the above-mentioned crossover, the middle the so-called
``waterfall,'' not of interest here, and the highest an ``anomalous
enhancement'' of the LDA-predicted band-width, by about 0.8~eV visible below
$\varepsilon_F$ in optimally doped BSCCO, i.e.\ roughly twice that in full. In
principle, LDA calculations are able to reach lower energy scales than small
clusters. Thus no ab initio effort has so far reached the vicinity of the
Fermi level in high-T$_c$ superconductors. Below the crossover scale, the
physical regime of the electrons actually undergoing the transition to
superconductivity must still be inferred from phenomenological analysis of
low-energy experiments.

The optimally doped region is equally well connected to the metallic overdoped
region as to the underdoped one. Attempts to understand the high-T$_c$
perovskites at optimal doping starting from the paramagnetic metal have a long
standing on the hole-doped side~\cite{Kotliar88}. They have mostly been framed
in terms of the mean-field slave-boson (MFSB) formalism, allowing for
effective multiband approaches, in which the values of the orbital parameters
depart significantly from those expected on the basis of high-energy
experiments and LDA calculations~\cite{Qimiao93}.

Instead of taking the high-energy ``bare'' regime as the starting point, and
introducing many-body effects to reach lower energies, MFSB phenomenology goes
in the opposite direction. Beginning with unprejudiced zeroth-order fits at
the Fermi level, one introduces the minimal perturbation necessary to model
the lowest crossover. Such an analysis of angle-resloved photoemission
(ARPES)~\cite{Sunko05-1} has recently been extended to electron-doped
compounds~\cite{Sunko07}, and produced the interesting result, that their
lowest-energy sector is in a very similar physical regime as in the hole-doped
ones~\cite{Sunko05-1,Mrkonjic03}: a large direct oxygen-oxygen hopping
$t_{pp}$, significantly larger than the effective copper-oxygen hopping
$t_{pd}$, and with a band-width of the same order as the effective
copper-oxygen splitting $\Delta_{pd}>0$, in the hole picture:
\begin{equation}
U_d\gg4|t_{pp}|\,
\raisebox{-1mm}{$\displaystyle\stackrel{\displaystyle >}{\sim}$}
\,\varepsilon_p-\varepsilon_d\equiv
\Delta_{pd}\gg t_{pd}^2/\Delta_{pd}.
\label{phr}
\end{equation}
The ARPES scales on which this observation is based are 0.2~eV in BSCCO and
0.5~eV in NCCO. The lowest-energy crossover was modelled by a $(\pi,\pi)$
boson, whose scales were associated with AF in the same compounds. The
low-energy regime (\ref{phr}), markedly different from (\ref{qcr}), is
characterized in particular by an ``anticrossing'' of the wide oxygen band
with the copper level, so that the Fermi level is found in an
oxygen-dominated, strongly dispersive part of the Brillouin zone, even though
$\varepsilon_d<\varepsilon_p$. As will now be shown, a directly analogous
regime exists in the ionic limit.

\section{Model and technique}

\begin{figure}
\includegraphics[height=50mm]{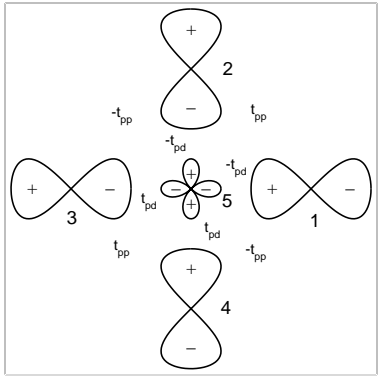}
\caption{Orbital parameters and symmetries used in the article. In this sign
convention $t_{pp}<0$ is physical~\cite{Golosov98}.}
\label{figorb}
\end{figure}
The Hamiltonian of a single CuO$_4$ molecule is
\begin{eqnarray}
H&=&\varepsilon_d\sum_{\sigma}\hat{n}_{d\sigma}+
\varepsilon_p\sum_{\sigma,i=1}^4\hat{n}_{p,i\sigma}
\nonumber\\\nonumber &&+
t_{pd}\sum_{\sigma}\left(\ad\Pl+\Pd\al\right)+
U_d\hat{n}_{d\uparrow}\hat{n}_{d\downarrow}\\\nonumber &&+
t_{pp}\sum_{\sigma}
\left(\pdx\py+\pdmx\pmy+\mathit{c.c.}\right)\\\label{ham} &&-
t_{pp}\sum_{\sigma}
\left(\pdmx\py+\pdmy\px+\mathit{c.c.}\right),
\end{eqnarray}
where $\Pl=\pmx+\pmy-\px-\py$ describes the four oxygens around the copper,
and $\hat{n}_{d\sigma}$, $\hat{n}_{p,i\sigma}$ are the number operators of the
copper and oxygen orbitals. The relevant orbitals and
parameters~\cite{Golosov98} are shown in Figure~\ref{figorb}. All 25 states of
two holes with opposite spins on the five orbitals are taken into account.

Standard symmetry analysis is used to reduce the Hamiltonian (\ref{ham}) to
block-diagonal form. The basic symmetry is of the square, corresponding to the
group $D_4$. Take the $z$-axis, orthogonal to the molecular plane, to be the
spin quantization axis. Since the spins have total projection zero, flipping
all spins is a symmetry operation, and since all orbitals are planar, this is
equivalent to a reflection in the plane. Then $D_{4h}$ is the magnetic (color)
group of the problem, and both the Zhang-Rice singlet, $\left|ZR\right>$, and
the doubly-occupied copper state, $\left|d^8\right>$, fall in its $A_{1u}$
representation, containing five singlets.

As usual, the main symmetry group does not fix the state vectors in full. It
is easy to complete the reduction by hand, but instructive to identify the
reasons for the remaining ambiguities. One is, as expected, symmetry with
respect to a subgroup, $D_{2h}\subset D_{4h}$. The particular $D_{2h}$ group
involved contains the spin-flip and three $180^\circ$ rotations, about the
$z$-axis and about the two diagonals at $\pm 45^\circ$ relative to the Cu--O
axes. It stands out because its operations change the sign of the
copper-oxygen overlap $t_{pd}$, while the energies depend only on $t_{pd}^2$.

The more interesting reason is dynamical symmetry, imposed by the assumed
degeneracy of the four oxygen orbitals. It eventually affects only two
$B_{1u}$ vectors, but to fix them ``honestly,'' the full dynamical symmetry
apparatus must be invoked, even on such a small problem as the present one.
The group elements are the Hamiltonian terms which break the oxygen
degeneracy. Two are visible in (\ref{ham}), proportional to $+t_{pp}$ and
$-t_{pp}$, respectively. If these are represented by matrices in one-particle
space, multiplying them generates the third, which corresponds physically to
next-nearest-neighbor (n-n-n) hops from one oxygen to the other on the same
axis. These three (and the identity) form the Abelian group $D_2$, appearing
here as a dynamical symmetry group. The n-n-n term must be diagonalized
simultaneously with the sum of $t_{pp}$ terms in (\ref{ham}) to separate out
the last vector without manual intervention.

The outstanding physical feature of dynamical symmetries is that they can
connect states with doubly and singly occupied degenerate sites, obviously
impossible by geometrical transformations. Two technical advantages of
dynamical over geometrical symmetry groups are that they grow with the size of
the problem, and depend on topology rather than shape. Both are significant
when the clusters are motivated by solid-state investigations, and contain
many equivalent sites. E.g., the dynamical group of the four oxygens in
two-particle space is $Z_2^4$, with sixteen elements, again Abelian. It is not
needed here because the geometrical symmetry already reduces the problem to
very small blocks. On the other hand, as a molecule becomes larger, the blocks
due to dynamical symmetry are expected to become smaller than those selected
by geometrical symmetry, especially if the molecule is of odd shape, as is
common in Monte Carlo calculations.

\section{Result}

The complete reduction gives ten stationary states unaffected by $t_{pd}$. In
addition the Hamiltonian contains six 2$\times$2 blocks and one 3$\times$3
block. The latter is found in the $A_{1u}$ subspace, connecting the states
$\left|d^8\right>$, $\left|ZR\right>$, and an oxygen two-hole singlet,
$\left|p^{-2}\right>$: 
\begin{eqnarray}
\left|p^{-2}\right>&=&\frac{1}{4}\left(
\updo{1} + \updo{2} + \updo{3} + \updo{4}\right.\\\nonumber
&&\left.+ \updw{1}{2} - \updw{1}{3} - \updw{1}{4}
 - \updw{2}{3} - \updw{2}{4} + \updw{3}{4}
\right).
\label{p2}
\end{eqnarray}
where $\updo{i}=\updwa{i}{i}$ are doubly occupied states, and $\updw{i}{j}=
\updwa{i}{j}-\dwupa{i}{j}$ are antisymmetrized singly occupied states.
The numbers 1--4 refer to the four oxygens, as in Fig.~\ref{figorb}.
The block reads:
\begin{equation}
\hspace{-5mm}\parbox[b]{1.4cm}{
$
\begin{array}{r}
\left|d^8\right>\,:\\
\left|ZR\right>\,:\\
\left|p^{-2}\right>\,:
\end{array}
$
}
\parbox[b]{6.2cm}{
$
\left(
\begin{array}{ccc}
2{\varepsilon_d}+U_d & 2 \sqrt{2}{t_{pd}} & 0 \\
2 \sqrt{2}{t_{pd}} & {\varepsilon_d}+{\varepsilon_p}+2{t_{pp}} 
& 2 \sqrt{2}{t_{pd}} \\
0 & 2 \sqrt{2}{t_{pd}} &2{\varepsilon_p}+4{t_{pp}}
\end{array}
\right)
$
}
\label{model}
\end{equation}
The states $\left|p^{-2}\right>$  and $\left|d^8\right>$ have an equally large
coherence factor $2 \sqrt{2}{t_{pd}}$ with respect to $\left|ZR\right>$, as
found by perturbation theory~\cite{Zhang88}. However, the coherence factor of
the oxygen-oxygen overlap $t_{pp}$ is twice as large for $\left|p^{-2}\right>$
as for $\left|ZR\right>$. Since the physical sign of $t_{pp}$ is negative,
this is a generic mechanism to destabilize the ZRS for all values of $U_d$
(including $U_d\to\infty$), which identifies $\left|p^{-2}\right>$ as a
relevant perturbation of the Zhang-Rice ground state.

By the same analysis, the orbital-antisymmetric singlet $\left|OA\right>$,
also considered by Zhang and Rice, appears in a 2$\times$2, $B_{1u}$ block
with another oxygen singlet, call it $\left|pp\right>$:
\begin{equation}
\parbox[b]{1.4cm}{
$
\begin{array}{r}
\left|OA\right>\,:\\
\left|pp\right>\,:
\end{array}
$
}
\parbox[b]{4cm}{
$
\left(
\begin{array}{cc}
{\varepsilon_d}+{\varepsilon_p}-2{t_{pp}} 
& 2 t_{pd} \\
2 t_{pd} &2{\varepsilon_p}
\end{array}
\right)
$
}
\label{oa}
\end{equation}
The splitting here is also $\Delta_{pd}+2t_{pp}$, as between
$\left|p^{-2}\right>$ and $\left|ZR\right>$ in (\ref{model}), but it occurs
because the copper-oxygen singlet is antibonding ($t_{pp}<0$), while the
oxygen singlet is non-bonding with respect to $t_{pp}$. Its structure is
\begin{equation}
\left|pp\right>=\frac{1}{2\sqrt{2}}\left(
-\updo{1} + \updo{2} - \updo{3} + \updo{4}+\updw{1}{3} - \updw{2}{4}
\right).
\end{equation}
The orbital-antisymmetric state remains irrelevant in the presence of
oxygen-oxygen hopping. The $t_{pd}$-coherence factors in (\ref{model}) and
(\ref{oa}) may be identified by perturbation theory~\cite{Zhang88}, without
the need to distinguish between $\left|p^{-2}\right>$ and $\left|pp\right>$.
However, once $t_{pp}$ is introduced, the different effects on
$\left|ZR\right>$ and $\left|OA\right>$ are related to the different symmetry
properties of $\left|p^{-2}\right>$ and $\left|pp\right>$. The Zhang-Rice
singlet $\left|ZR\right>$ is destabilized because $\left|p^{-2}\right>$ has a
large coherence factor $4t_{pp}$, while $\left|OA\right>$ moves upwards ``out
of the way,'' and $\left|pp\right>$ cannot influence the ground state because
it is unaffected by $t_{pp}$ (but of course it can anticross with
$\left|OA\right>$ as the latter rises).

\begin{figure}
\includegraphics[height=50mm]{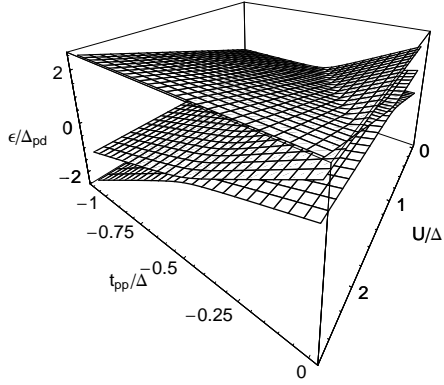}
\caption{Parameter space of the model (\ref{model}).}
\label{figpar}
\end{figure}
The coherence factors in the reduced model (\ref{model}) are the largest in
the whole space, so for $|t_{pd}|>0$ and $t_{pp}<0$ it contains the ground
state. The relevant parameters are $U_d/\Delta_{pd}$ and $t_{pp}/\Delta_{pd}$.
The precise value of $t_{pd}$ is irrelevant, as long as one is in the
parameter regime (\ref{phr}), corresponding to optimal doping. Choosing it low
for graphical reasons, one obtains Figure~\ref{figpar}. At the origin
($U_d=0$, $t_{pp}=0$) is the weak-coupling regime, where the ground state is
dominated by $\left|d^8\right>$. As $U_d$ increases, this state goes through
two anticrossings with the others, ending up highest in the Zhang-Rice regime
($U_d$ large, $|t_{pp}|$ small), where $\left|ZR\right>$ is the ground state.
Now as $|t_{pp}|$ increases, the state $\left|p^{-2}\right>$ goes down at
twice the rate of $\left|ZR\right>$, so it must eventually anticross with it.
Thus when both $U_d$ and $|t_{pp}|$ are large, the ZRS is no longer the ground
state. 

\section{Discussion}

The present paper describes a simple and robust symmetry-based mechanism which
prevents the isolation of the ZRS as the only candidate for the ground state
of the CuO$_4$ ion when $U_d$ is large. To wit, if the oxygen-oxygen hopping
$t_{pp}$ is significant, the local environment will be in a charge transfer
regime involving two states, the ZRS and an oxygen singlet. The two are mixed
by the symmetry-enhanced overlap $2 \sqrt{2}{t_{pd}}$ across the
symmetry-reduced gap $\Delta_{pd}+2t_{pp}$. This is the metallic counterpart
to the result~\cite{Eskes93} that the antiferromagnetic interaction $J$ near
half-filling is also dominated by the charge-transfer term
$t_{pd}^4/\Delta_{pd}^3$, while the Mott-Anderson contribution $t_{pd}^2/U$ is
negligible.

It has long been established that ARPES fits require significant ratios of
$|t_{pp}/t_{pd}|$ in order to agree with the data~\cite{Qimiao93}. Recently
there appeared detailed ARPES fits, both qualitatively and quantitatively in
accord with experiment, in practically the same orbital regime for
NCCO~\cite{Sunko07} and BSCCO~\cite{Sunko05-1}, which established the
anticrossing regime (\ref{phr}) as an appropriate phenomenology for these
compounds, meaning the ratio $|t_{pp}|/\Delta_{pd}$ is also enhanced,
especially so on the electron-doped side.

Changes in orbital parameters as a function of doping are a standard feature
of effective theories such as mean-field slave-boson (MFSB), which account for
the effects of strong correlations by changing $t_{pd}$ and $\Delta_{pd}$. The
question, as always with effective band models, is what this means for the
local environment: do local hops ``see'' the bare or the renormalized band
parameters. The conclusion that the effective band regime at optimal doping is
reflected in an analogous local regime, which implies the destabilization of
the Zhang-Rice singlet by the above symmetry mechanism, is supported by two
independent arguments, one phenomenological and one theoretical.

The recently emphasized~\cite{Meevasana07} failure of LDA calculations to
reproduce experimentally observed band widths is phenomenologically
significant, because LDA predicts a smaller width than is actually observed,
so any conceivable inclusion of additional correlations would only make the
discrepancy larger. Notably, LDA calculations agree with high-energy
experiments and calculations in the ionic limit. The anomalous enhancement of
hopping via the oxygen sites is consistent by symmetry with the observed
discrepancy, which is large in the nodal direction, where the oxygen band is
strongly dispersive. The enhancement of the ARPES bandwidth of 1--2~eV over
LDA predictions~\cite{Meevasana07} is quantitatively in good agreement with
the value of $t_{pp}$ needed to enter the anticrossing regime, i.e.\
destabilize the ZRS.

Theoretically, the parameter space of the three-band MFSB model has been
extensively investigated~\cite{Mrkonjic03} from the point of view of outcomes,
namely: which regions of renormalized orbital parameters are indicative of
strong renormalization of input parameters by the MFSB mechanism, and which
imply preservation of the bare parameters. In the so-called ``resonant-band''
regime, characterized by a narrow effective bonding band and copper-dominated
wave functions, the quasiparticle weight in the conduction band is
significantly reduced, being proportional to the doping $\delta$ instead of
the total hole content $1+\delta$. In that case, the band calculations cannot
be directly related to a local picture, since the doped-hole weight is locally
either one or zero. However, in the anticrossing regime (\ref{phr}), the
opposite occurs: MFSB theories with Gaussian fluctuations depart from the
resonant-band regime and acquire a full quasiparticle weight in the Brillouin
zone, only with a reduced $t_{pd}$~\cite{Mrkonjic03}. They are therefore
comparable and consistent with the present ionic ones, as long as both are in
the anticrossing limit. Thus the present work reaffirms the
possibility~\cite{Abrikosov89} that the optimizations leading to a large
direct oxygen-oxygen overlap below the crossover scale occur already at the
local orbital level.

Realistically, $t_{pd}$ is not so small, so the ground state in real materials
retains substantial copper content. In the local picture, this means a
significant ZRS content in the ground state. However, even if $t_{pd}$ is as
large as $|t_{pp}|$, giving a large anticrossing gap $\sim\Delta_{pd}$, the
ground state still contains nearly equal components of $\left|ZR\right>$ and
$\left|p^{-2}\right>$. The system is physically in the charge-transfer regime,
which also provides the dominant contribution~\cite{Eskes93} to $J$. In that
case, the reduction to one band by the limit~\cite{Zhang88}
$\Delta_{pd}=U_d-\Delta_{pd}\to\infty$ is purely formal, since that limit
replaces the charge-transfer regime \emph{a priori} by the Mott regime.

While the MFSB calculations allow both for regimes with strong renormalization
and no renormalization~\cite{Mrkonjic03}, the derivation of the $t$--$J$ model
from the ZRS~\cite{Zhang88} implies by construction that the band parameters
are locally relevant. However, it has been noticed~\cite{Macridin05} that in
order to face experiment, the $t$--$J$ model needs to be supplemented by
additional next- and next-next-nearest neighbor hops, making up the so-called
$t$--$t'$--$t''$--$J$ model. Since the $t'$ term breaks particle-hole symmetry
in the same way as the $t_{pp}$ term in the three-band model, its appearance
in practical calculations~\cite{Kyung04} is immediate evidence that the
oxygen degree of freedom cannot be eliminated from the physical picture near
optimal doping. The need to introduce $t'$ at the band level translates
microscopically into the need to admix the oxygen singlet (\ref{p2}) to the
ZRS, to describe the local electronic environment around optimal doping. (When
the oxygen is treated explicitly in the band picture, $t_{pp}$ has the same
sign for both n- and p-doping~\cite{Sunko05-1,Sunko07}, as also found in LDA
calculations~\cite{Korshunov05}, consistently with its simple chemical
interpretation.)

To see the real effect of the $t'$ term, the dispersion of the $t$ and $t'$
terms may be rewritten
\begin{eqnarray}
\lefteqn{-2t(\cos k_x+\cos k_y)
-4t'\cos k_x\cos k_y=}&&\nonumber\\
&&-4(t+t')+4(t+2t')(\sin^2\frac{k_x}{2}+\sin^2\frac{k_y}{2})
-16t'\sin^2\frac{k_x}{2}\sin^2\frac{k_y}{2},
\end{eqnarray}
separating contributions by particle-hole symmetry. The second term is
manifestly symmetric around half-filling, while the third is not, since it
achieves its minimum value (zero) along the non-dispersive lines $k_x=0$ and
$k_y=0$. In practice $tt'<0$, and the ratio $|t'/t|\sim 0.2$--$0.4$. For
$t'=-0.3 t$, a common enough parametrization around optimal doping,
$|4t'/(t+2t')|=3$, so the symmetry-breaking term enters with \emph{three
times} the contribution of the symmetry-preserving term. Thus the apparently
small values of $|t'/t|$ found in the literature are quite misleading in terms
of the actual amount of particle-hole symmetry breaking needed to achieve
realistic Fermi surface fits within the $t$--$t'$--$J$ model. In particular,
for $t'=-t/2$ the symmetry-preserving term vanishes, so the $t$--$t'$--$J$
model passes through a quantum phase transition in the vicinity of its
applicable parameter range.

By symmetry, this is the same type of transition appearing in the present
paper as the destabilization of the ZRS in the three-band case. Physically,
they are different insofar as the three-band model destabilizes the ZRS by
oxygen hopping in low order, while in the $t$--$t'$--$J$ model the Zhang-Rice
singlets are preserved by assumption, and the transition occurs by the
additional $t'$ overlap among them, i.e.\ in high order in the original
three-band model. But to account for ARPES data, both models agree that such a
transition is approached near optimal doping.

The limitation of the above arguments to the metallic phase should be kept in
mind. Near half-filling, the oxygen hopping is effectively reduced by the
antiferromagnetic interactions, and precisely at half-filling, the system is
insulating independently of the value of $t_{pp}$. In the insulating phase,
the high-energy ionic limit is recovered, both theoretically and
experimentally. As $t_{pp}$ becomes more effective with doping, locally the
first admixture to the ZRS appears, identified here as the oxygen singlet
(\ref{p2}). Thus a broad picture of the optimal-doping regimes of high-T$_c$
compounds emerges, in which NCCO is deep in the anticrossing regime,
$4|t_{pp}|\gg \Delta_{pd}$, and BSCCO is at its limit, $4|t_{pp}|\approx
\Delta_{pd}$. On the other hand, the remarkable Fermi-surface evolution of
LSCO with doping~\cite{Ino99} is indicative of strong renormalizations of the
band parameters, especially when the system is underdoped. In such a
resonant-band regime, it was also shown directly~\cite{Jankovic00} that the
strong doping-dependence of $t_{pp}$ in the MFSB fits did not imply a similar
variation of the local parameters.

The present analysis points to one pitfall of considering the oxygen part of
the ZRS as a single-particle state in its own right, as if the
hole on the copper were unhybridized with the oxygens. However, the same
caution applies to the oxygen singlet found here. In particular, one should be
careful in interpreting the corresponding Bloch states. The propagation of
strongly correlated cluster states in a lattice is at present largely
unexplored in more than one dimension, with controlled analytic results
available only in special limits, which themselves amount to uncontrolled
simplifications. Such is the limit of infinite dimensions~\cite{Georges96},
physically corresponding to diffusion, or the Falicov-Kimball
limit~\cite{Sunko05}, which is dispersive, but replaces dynamical constraints
by geometrical ones.

The model presented here is as sparse as possible, allowing for the
experimentally observed regime with only three state vectors. To avoid being
too minimal, some attention should be paid to the neglected Coulomb integrals
$U_{p}$, $V_{pp}$ and $V_{pd}$, the on-site oxygen, intracell oxygen, and
copper-oxygen repulsions, respectively. They do not break the symmetries
above, introducing only matrix elements within each symmetry block. In
particular, they connect the ground-state block (\ref{model}) with the other
two $A_{1u}$ states by matrix elements $U_{p}/4$, smaller than the terms
$2\sqrt{2}t_{pd}$. On the diagonal of this block, the Coulomb terms can be
reabsorbed in the site energies and copper on-site repulsion $U_d$. For
realistic $V_{pp}\ll U_{p}$, this leads to a small reduction in the
copper-oxygen splitting $\Delta_{pd}$, and practically no change in $U_d$,
while for large $V_{pp}=U_{p}/2$, a useful limiting case, the effective $U_d$
is increased and the effective $\Delta_{pd}$ decreased, both by about 1~eV.
Thus the inclusion of neglected integrals mildly tends to push the system
further into the anticrossing, $U_d\to\infty$ regime.

An extension of the Zhang-Rice argument to the electron-doped side is only
possible with larger molecules, containing more than one copper atom. This
would allow to study the competition of magnetic and charge ordering at the
local level. It also opens a way to compare the Zhang-Rice and
Emery-Reiter~\cite{Emery88} constructions directly on the hole-doped side.
Obviously, the spaces involved are much larger. It is on these larger
problems, both theoretical and numerical, that hope is invested in the
assignment of quantum numbers by dynamical symmetries, as introduced in the
present work.

\section{Conclusion}

It has been found that in the presence of local repulsion on copper, the
direct oxygen-oxygen overlap is a relevant perturbation of the Zhang-Rice
singlet. With the possible exception of LSCO, real hole- and electron-doped
high-T$_c$ superconductors appear to be in a physical regime where it is
strongly mixed with a pure oxygen singlet.

\section{Acknowledgments}
Conversations with S. Bari\v{s}i\'c are gratefully acknowledged. This work was
supported by the Croatian Government under Project~{119-1191458-0512}.

\end{document}